\documentclass[3p,review, 12pt, authoryear,preprint,times]{elsarticle}

\usepackage{amssymb,amsthm,amsmath,amsfonts}

\usepackage[latin1]{inputenc}

\usepackage{todonotes}
\usepackage{csquotes}
\usepackage{breakcites}
\usepackage{graphicx}
\usepackage{color, natbib}
\usepackage{placeins}
\usepackage{changes}
\usepackage{appendix}
\usepackage{enumitem}
\usepackage{tikz}
\usepackage{hyperref} 
\usepackage{soul}
\usetikzlibrary{positioning}
\usetikzlibrary{shapes}
\usetikzlibrary{patterns}
\tikzset{
	dots size/.store in=\dotssize,
	dots size=1pt,
	dots spread/.store in=\dotsspread,
	dots spread=10pt
}
\makeatletter
\pgfdeclarepatternformonly[\dotssize,\dotsspread]{mydots}
{\pgfpointorigin}
{\pgfpoint{\dotsspread}{\dotsspread}}
{\pgfpoint{\dotsspread}{\dotsspread}}
{
	\pgfsetcolor{\tikz@pattern@color}
	\pgfpathcircle{\pgfpoint{\dotsspread/2}{\dotsspread/2}}{\dotssize}
	\pgfusepath{fill}
}
\makeatother

\usepackage[figuresright]{rotating}

\newtheorem{theorem}{Theorem}
\newtheorem{lemma}{Lemma}
\newtheorem{definition}{Definition}
\newtheorem{proposition}{Proposition}
\newtheorem{corollary}{Corollary}

\begin{document}

\begin{frontmatter}


\title{Auctioning Corporate Bonds:
	A Uniform-Price under Investment Mandates\tnoteref{label1}}
\author{Labrini Zarpala}
 \ead{l.zarpala@uu.nl}
\fntext[]{I am genuinely grateful to my Ph.D. supervisor Dimitris Voliotis for his guidance and support and to Nikolaos Egglezos. Moreover, I thank the Royal Economic Society for the Mentoring Initiative 
and Adopt Paper Initiative 
and the participants of the 30th Annual Conference of the European Financial Management Association, Conference on Mechanism and Institutional Design, and the 88th International Atlantic Economic Conference for the valuable comments. Last, I am thankful to the journal Games for the grant awarded.}
\address{ Utrecht School of Economics, Kriekenpitplein 21-22, 3584 EC Utrecht, The Netherlands }





\begin{abstract}

This paper examines how risk and budget limits on investment mandates affect the bidding strategy in a uniform-price auction for issuing corporate bonds. I prove the existence of symmetric Bayesian Nash equilibrium and explore how the risk limits imposed on the mandate may mitigate severe underpricing, as the symmetric equilibrium's yield positively relates to the risk limit. Investment mandates with low-risk acceptance inversely affect the equilibrium bid. The equilibrium bid provides insights into the optimal mechanism for pricing corporate bonds conveying information about the bond's valuation, market power, and the number of bidders. These findings contribute to auction theory and have implications for empirical research in the corporate bond market.


\end{abstract}

\begin{keyword}
	Bond Market \sep Auctions \sep Investment Mandates \sep Market Design


\end{keyword}

\end{frontmatter}


\section{Introduction}
This article investigates the application of a uniform price auction for the pricing of corporate bonds in the primary market. 
Investors compete by submitting bids entailing information about how much money they want to invest for a specific interest rate. The bond's allocation results from an ascending ranking  of interest rates till the supply side (amount to be raised by the issuer) is exhausted. All investors buy a portion of the issuance at the same  \enquote{market clearing} (or stop-out) yield. Demand below this yield is fully allocated, while when there is a tie at the stop-out yield, the marginal demand is prorated.

The bidding structure indicates that risk considerations and budget constraints are essential in shaping bidding strategies. The term investment mandate includes those two factors and describes a set of guidelines that asset managers must follow when making investments on behalf of the investor. Usually, it foresees a certain amount of money invested in the primary market irrevocably committed to investment-graded bond strategies. 

Last decade, the investment mandates for corporate bonds have been in the spotlight and made investors feel more comfortable with their allocations. For example, the prospectus of Invesco Global Investment Grade Corporate Bond Fund\footnote{https://www.invesco.nl/dam/jcr:cb6153ab-0bd2-47fd-$80da-8d96921bd6f3/LU1075208725_EN_NL.pdf$} disseminated to potential clients explicitly describes the fund's mandate to invest a portion its total assets in investment-grade corporate bonds specifying that it has a \enquote{\textit{limited ability to own high yield securities (rated no lower than BB at purchase) because it predominately uses the flexibility to invest in subordinated bonds\footnote{Subordinated bonds are unsecured and therefore riskier. Subordinated bondholders may receive a higher interest rate to offset possible losses. Banks often issue subordinated bonds to meet Tier II capital requirements rather than for financing purposes, and their issuance is sometimes cheaper.} issued by investment-grade companies rather than buying riskier assets.}} In other words, the asset manager's mandate is to invest only in investment-grade bonds in the primary market, as there is always the risk of the bond being downgraded after the issuance.

I conceptualize this critical feature of the bidding strategy for investment-grade bonds with the common knowledge assumption.  I assume that all asset managers know the supremum risk limit of all asset managers, i.e., to invest in an investment-grade issuance. This information is available ex-ante for all auction participants and is released by the credit rating agencies that signal the quality of the issuer. Due to the common knowledge assumption about the asset manager's beliefs for the risk limit (i.e., all participate in an investment grade issuance), I reduce the dimensionality of the asset manager's type only to the truthful announcement of the budget limit in a direct revelation mechanism as the risk limit is symmetric for all auction participants \citep{Hafalir2012, Dobzinski2012}. This mechanism uses a linear rule to allocate the bond and define the stop-out yield. 

To better adapt the model to the real-world corporate bond market, I include in the analysis the expectation that the asset managers have for the secondary market. The difference between the issuance yield (purchase price) and the expected yield in the secondary market (sale price) is the common value component of the auction. Even though this component is a common ex-post value \citep{Milgrom1982} for all asset managers, ex-ante is unknown. Each asset manager might have different information\footnote{The signals influencing investors' valuation can be classified but not limited to multiple corporate bond yield predictors, historical yield averages, and past returns \citep{Guo2021}.} about the value of the bond drawn from a signal space and a common distribution.

Additionally, I measure the market power over the issuance's yield with a sensitivity factor and use a benchmark interest rate associated with substantial risk bonds (\enquote{high-yield} bonds)\footnote{High-yield bonds, usually called \enquote{junk bonds}, offer a higher interest rate because of their higher risk of default. Companies with a higher risk of default issue these types of bonds. They offer a higher yield to attract more investors to compensate them for the additional undertaking risk.} as a measurement of fire selling risk\footnote{If the issuer is abruptly downgraded from BBB to junk status as a result of economic weakness, the fund with investment grade mandates could be forced to unload large amounts of bonds quickly \citep{BIS2019}.}. In the analysis, it is necessary to include the interest rate of \enquote{high-yield} bonds to approximate asset managers' truthful valuation during the auction. The interest rate of \enquote{high-yield} bonds is the lowest price for selling the bond in the secondary market and estimates potential losses relative to the anticipated selling price.

To interpret the results, I decompose the equilibrium bid into components. Each component has relative importance in the equilibrium affected by the weight of the infimum allocation over the symmetrical, assuming that the mechanism allocates a quota in the issuance equal to the reported budget. The first weighted component of the equilibrium bidding strategy is the infimum bid, which reflects the risk limit imposed on the mandate. Ceteris paribus, the infimum  bid has a positive relationship with the equilibrium strategy and an inverse relationship with the supremum risk limit.

The equilibrium strategy's second weighted component measures: the asset manager's oligopolistic market power, the spread between the expected yield in the secondary market and the yield of \enquote{high-yield} bonds, and the number of bidders. All other things being equal, bidders (asset managers) with symmetrical strong market power, i.e., investment mandates with low-risk acceptance, inversely affect the equilibrium bid. This is similar to the Cournot oligopoly.  Notwithstanding that the market power arises endogenously in the  equilibrium bid, like the existing theory \citep{Back1993,Quarterly2014}, I show that the underpricing might be less severe due to the risk limitations imposed in the mandate. This condition holds for downwardly bounded continuous bidding strategies, while the theory proves it under discrete bids \citep{Kremer2004a}. 

Another similarity with a symmetrical Cournot oligopoly is that the number of bidders inversely affects the equilibrium bidding strategy. Still, the equilibrium yield of the issuance remains unaffected by the number of bidders. The spread between the expectation for the secondary market and \enquote{high-yield} bonds captures the potential loss on the value of the bond, and it positively affects the equilibrium bid. If bidders anticipate lower yields (higher prices) in the secondary market, they will increase their demand over the issuance; all other things being equal. If bidders expect a yield close to the substantial risk bonds in the secondary market, then the equilibrium bidding strategy is restricted to the first weighted component (infimum bid).

The contribution of this paper is twofold: from an auction theoretical perspective, I contribute to the discussion around uniform auctions by exploiting how an exogenous risk limit and a budget limit can be embedded in the bidding strategy. I construct a model where budget limits are truthfully reported in the auction mechanism, and the risk limit is common knowledge. I result in a non-unique symmetric Bayesian-Nash equilibrium and explore how risk limit bound the bidding strategy. From the financial literature perspective, I contribute to discussing the optimal mechanism for pricing corporate bonds. To the best of my knowledge, this paper is the first to attempt that proves an equilibrium bid for the pricing of corporate bonds in the primary market. The equilibrium bid conveys information about the bond's valuation, the market power, and the number of bidders.  This result can be exploited for further empirical research. 

The extant literature has focused only on pricing government bonds, analyzing two auction formats: uniform or discriminatory\footnote{It is a sealed bid auction for homogenous items in which each winner pays an amount equal to the sum of the bids placed for the items allocated.}. However, neither empirical research \citep{Nyborg1996,Tenorio1997,Binmore2000}, nor auction theory \citep{Quarterly2014,Back1993,Wang2002,Bikhchandani1993, Ausubel2014}  offer a constraining reason for preferring discriminatory to uniform auctions. Another strand of the literature focuses on fixed price offerings versus uniform price auctions for IPOs \citep{BIAIS20029,SHERMAN2005615}. Experiments show that uniform price auctions are superior to fixed price offerings in raising revenues \citep{Zhang2009}.  

Exploiting auctions in issuing corporate bonds is essential, as more attention has yet to be given. The theory shows that auctions outperform post-pricing selling \citep{10.2307/136330}. Post-price selling is the current mechanism applied to issuing corporate bonds, called book-building\footnote{An underwriter (investment bank) undertakes the competitive sale and the efficient allocation of the new issuance. The underwriter markets the offering to investors, asking for a non-binding Indication of Interest. This pre-market information is less costly for the underwriter to measure the demand and to adjust the offering's price and coupon if needed \citep{Iannotta2010a, Nikolova2020, Habib2007}}, and has certain caveats, such as allocation weakness and mispricing over new issuances \citep{Bessembinder2020, Cornelli2001,JENKINSON2004}. Therefore, I focus only on uniform auctions for the primary market of corporate bonds and examine the existence of a symmetric equilibrium bid bounded by investment mandates' guidelines.

The paper is organized as follows. The following section \ref{Sec:Model} contains a formal analysis and describes the model as a direct revelation mechanism. In the same section, I introduce the concept of risk limit. Section \ref{Sec:Eq}, studies bidders' incentives and provides the proof of a Bayes-Nash symmetric equilibrium for independent signals performing the respective comparative statics. The last sections, \ref{discussion} and \ref{Concl}, discuss the previous sections' outcomes and conclude. Some proofs are expanded on Appendix A.

\section{Model}\label{Sec:Model}
\subsection{Preliminaries} \label{Preliminaries}
I assume a common value auction for the sale of a single unit of perfectly divisible bond 
with a face value equal to one, and $n$ competitive bidders, defined as a finite set $\mathcal{I}$=$\left\{{1,2, 3\dots n}\right\}$, with $n\geq3$. All bidders are risk-neutral and none of them is eligible to bid for the full face value of the bond. 

Each bidder $i$ has an upper-bound bidding stipulated by the investment mandate, defined as the \textit{budget limit} $c_i\in[\underline{c},\bar{c}]$,  as well as a \textit{risk limit}  $\mathit{r_i^\ell}\in[\underline{r},\bar{r}]$. The type of $i$ is defined as $\tau_{i}=(c_i,r_i^\ell)$, with $\tau \in \mathcal{T}$, which attributes bidders' preferences $\mathcal{T}:=[\underline{c},\bar{c}] \times[\underline{r},\bar{r}]$ to the eligible real intervals. Each type  $\tau_{i}$, is i.i.d. to a continuous joint cumulative function $F(\tau)= F_{c r^\ell}(c,r^\ell)$ commonly known to all bidders, and
fully supported by $f(\tau)>0$.

Ex-ante, the equilibrium yield of the secondary market, is an unknown random variable, $r^{s}\in[\underline{r}$, $\bar{r}$], with a cumulative distribution $H(r^{s})$ that is common knowledge to all bidders. Also, it has full support  $h(r^{s})>0$, $\forall r^{s}\in[\underline{r}, \bar{r}]$.  The expectation over the secondary market is denoted as $\mathbb{E}[r^{s}]$.

All information for the bidding strategies $\tau_{-i}:=(\tau_1,\dots,\tau_{i-1},\tau_{i+1},\dots\tau_n)$ is summarized in a joint cumulative function $G(r^{s},\tau_{-i})= H(r^{s})\times F(\tau_{-i})$, fully supported by $g(r^{s},\tau_{-i})>0$.
Additionally, bidders receive an independent private signal $s$ about 
the value of the bond and other bidders' preferences. This information for each bidder $i$ is embedded in $s_i\in \mathcal{S}$, where $\mathcal{S}$ is the signal space with infinite elements that allow each bidder's value to be a general function of all the signals.

The strategy of each bidder $i$ is a bid schedule:  
\begin{equation}\label{mapping:1}
	b_i(\cdot,\cdot|\tau_i): [\underline{r},\bar{r}] \times \mathcal{S} \rightarrow[0,1)
\end{equation}
defined on the signals' space $\mathcal{S}$ while $[\underline{r}, \bar{r}] \in R^*_+$ is the domain of eligible yields set by the auctioneer. Each schedule specifies the quantity demanded at a specific interest rate $r$ based on the different realizations of a private signal $s_i$ for the value of the bond.
Bid schedules are assumed to be continuously differentiable, decreasing to the interest rate $r$, and an increasing continuous function in the budget limit $c$. 

The outcome of the issuance  $(\alpha,\hat{r})$ consists of two components: a payment rule $\mathit{\hat{r}}$, i.e. the stop-out yield and an allocation rule $\mathit{\alpha}\in [0,1)$. I define the payment rule $\mathit{\hat{r}}$, and the allocation in the next section (see Definition \eqref{definition1} and \eqref{definition3} respectively).


For a strategy profile $b(\cdot)$ the payoff function 
of a risk-neutral bidder $i$
given the observed signal $s_i \in S$ is:

\begin{equation} \label{eq:2}
	{\mathbb{E}}_{(r^{s},\tau)}[\pi_{i}(b|s_i)]= \mathbb{E}_{\tau|s_i}\bigg[\bigg(\hat{r}(b) - \mathbb{E}[r^{s}]\bigg) \alpha_i(b)\bigg].
\end{equation} 


\subsection{Market Mechanism}

In this section, I will elaborate on the mechanism that produces the outcomes of the auction. The process starts with the simultaneous submission of bids. 
Given Equation \ref{eq:2}, the intrinsic value of the bond is determined by the term $\hat{r} - \mathbb{E}[r^s]$. Bidders submit their bids based on the prior common distribution attributing beliefs about other bidders' types. 

I focus on a direct revelation mechanism. Bidders truthfully reveal their types $\tau_i(c_i, r_i^\ell)$ through their bidding strategy $b_i$ (mapping \ref{mapping:1}) in the mechanism. Since the second dimension of type, the $r^\ell$  is common knowledge and symmetrical for all bidders, the only dimension of the type that needs to be reported truthfully in the mechanism is the budget limit $c_i$.




After collecting all strategies with the truthfully reported types (i.e., budget limits $c_i$), the auction outcome follows the uniform pricing rule \citep{Wang2002, krishna2010}. The individual demands are rearranged in ascending order concerning the yield and aggregated, the \textit{stop-out yield} $\hat{r}$ of the auction is defined as the supremum yield, at which the aggregate excess demand is nonnegative, i.e., exceeds the face value of the bond, which is normalized to one. The stop-out yield is a market-clearing yield. The following linear rule in equation \eqref{eq:4} specifies the aggregate inverse demand function for the issuance and, in the symmetric case, applies to each bidder's $i$ inverse demand function for a bidding strategy $b_i(\cdot,\cdot|\tau_i)$.

\begin{definition}\label{definition1}
	All winning bidders receive a stop-out yield $\hat{r}$, after the auction ends such as  
	

\begin{equation}\label{eq:3}
		\hat{r} =sup\{r\in [\underline{r}, \bar{r}]\big| \displaystyle D(b)\geq 1\}
	\end{equation}
	and I assume that $\hat{r}$ follows a linear rule:
	\begin{equation}\label{eq:4}
\hat{r}=\begin{cases}
			\Theta-\theta D(b) &, \mbox{if $D(b)\geq 1$, with $b,\theta>0$ and  $\Theta>\theta D(b)$} 
\\ 0 &\mbox{, otherwise}
		\end{cases}
	\end{equation}
	\\	
	where 
	$D(b)=\sum\limits_{i=1}^{n}b_i(\cdot,\cdot|\tau_{i})$ for $n\geq2$, $\Theta$ is the yield of \enquote{high-risk} bonds and $\theta$ is an exogenous sensitivity factor of the stop-yield towards a change on $D(b)$.
	
\end{definition}

To simplify the analysis, I assume that the parameter $\theta$ remains symmetric for all bidders, which means that they all have the same market power over the yield's structure. The parameter $\Theta$ acts as a benchmark interest rate indicating the interest rate of \enquote{junk bonds}. During the auction, $b(\cdot,\cdot|\tau_{i})=0$ for an interest rate equal to $\Theta$. However, if the bond is downgraded after the auction to a \enquote{junk bond}, $\Theta$ will be the interest rate that bidders will fire sell their allocation despite their expectation for $r^s$.



Below, I define an allocation rule that specifies how the asset is allocated so that no bidder gets more than his bid. 


\begin{definition} \label{definition3}
	Under the absence of ties, the allocation rule is a one-to-one mapping from the set of bid schedules' profiles $b_i(\cdot)_{i=1}^{n}$ to non-negative allocations $\alpha_i\in(0,1)$, $\forall\mathit{i}\in\mathcal{I}$ with $\alpha(0)=0$, such that $D(b)=1$. For non-winners ${\alpha}_{j}=0$, $\forall\mathit{j}\in\mathcal{I}$. 

\end{definition}

The allocation that the mechanism generates is such that winners invest all their budget $c_i$ in the issuance.

\begin{lemma} \label{lemma_2.1}
	The auction admits to a direct revelation mechanism, where bidders truthfully report their budget limit.
\end{lemma}

\textit{Proof.} Suppose that bidder $i$ finds more profitable to report in the mechanism a budget $\tilde{c}$ than the truthfull budget $c^*$, with $\tilde{c} > c^*$. Then  the same bidder would have found it profitable to submit a bid $b(\tilde{c}) \geq b(c^*)$ which would result in $a(\tilde{c}) \geq a(c^*)$.
Then $\mathbb{E}[\tilde{\pi_i}] \geq \mathbb{E}[{\pi_i}^*]$  but since $b$ constitutes an equilibrium bid, this is impossible. The result is standard in
the literature \citep{Dobzinski2012,Hafalir2012}. \qedhere


\subsection{The concept of risk limits }\label{risklimits}

Now, I will explain how the risk limit affects the infimum bidding amount and the payoff of each asset manager. Let us assume a bidder $i$ who has to comply with an investment mandate with a supremum risk  $r_i^\ell \in[r^f,\bar{r}] $, where $r^f$ is the risk-free rate.  I further assume that bidder $i$ has a budget limit $c_i\in [0,\bar{c}]$, with $c_i^\ell$ to be the infimum bid associated with the risk limit $r_i^\ell$.


Figure 1 plots the inverse aggregate demand function of equation \ref{eq:4}. The intuition is that a highly demanded issuance will be priced at a low-interest rate. I plot
the asset managers' \enquote{bid} on the horizontal axis. Under the assumption that each asset manager reveals only one dimension of his type, i.e., the budget limit  $c_i $, the horizontal axis measures the amount that each asset manager $i$ is willing to bid. The vertical axis denotes the different \enquote {yield} levels associated with the bid. 
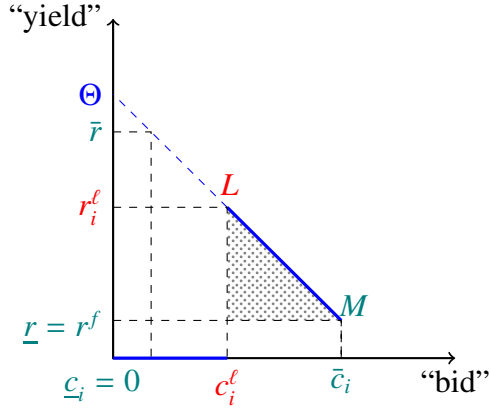
\begin{figure}
	\centering
	\begin{tikzpicture}
		
		\draw [<->,thick] (0,4.5) node (yaxis) [left] {\enquote{yield}}
		|- (4.5,0) node (xaxis) [below] {\enquote{bid}};
		
		\draw [very thick, blue] (0,0)--(1.5,0);
		\node [below left,teal] at (0.5,0) {$\underline{c}_i=0$};
		
		\node [below,red] at (1.5,0) {$c_i^{\ell}$};
		
		\node [left,red] at (0,2) {$r_i^{\ell}$};
		\node[left,red] at (1.8,2.3) {$\mathit{L}$};
		\node[left,teal] at (0,0.4) {$\underline{r}=r^{f}$};
		\node[below,teal] at (3,0) {$\bar{c}_i$};
		\node[left, teal] at (3.5, 0.7) {$M$};
		\node[left,teal] at (0,3) {$\bar{r}$};
		\node[left,blue ] at (0,3.5){$\Theta$};
		
		\draw [very thick, blue] (1.5,2)--(3,0.5); 
		\fill [pattern=crosshatch dots, opacity=0.7] (1.5,0.5) --(3, 0.5) --(1.5,2);
		
		\draw [dashed, blue] (1.5,2)--(0,3.5); 
		\draw [dashed] (0,2)--(1.5,2)--(1.5,0);
		\draw[dashed]  (0,3)--(0.5,3)--(0.5,0);
		\draw[dashed]  (0,0.5)--(3,0.5)--(3,0.5);
		\draw[dashed]  (3,0)--(3,0.5)--(3,0);

	\end{tikzpicture}		
	\centering
	\caption{Aggregate Inverse demand curve with risk and budget constraints.}
	\label{figure_2.1}

\end{figure}
\FloatBarrier

Bidders who participate in the auction will always demand a high stake at a low yield to maximize payoffs subject to the expectation for the secondary market. If all auction participants demand a high stake over the issuance, i.e., $c_i=\bar{c}$ the stop-out yield will be further reduced up to $\underline{r}=r^f$. However, at point M, bidders are indifferent to investing in issuance with a stop-out yield equal to the risk-free rate, so $b(\bar{c})=0$. In reverse, if all auction participants demand the infimum stake over the issuance, i.e., $c_i=c^\ell$, the stop out-yield will be set at the supremum risk limit $r^\ell$. 

From the individual demand perspective, the infimum demand of bidder $i$ is $c_i^\ell$, including an interest rate $r^\ell$ in his bidding schedule. However, a low bid at a high yield will not include him among the winners (uniform-price auction). His incentive is to exhaust budget $c_i$ in an investment-grade bond, requesting a relatively low yield that will ensure not only his winning but also a positive $\hat{r}-\mathbb{E}[r^s]$. Thus strategically, the bidder 
 will bid between $c_i^\ell$ and $\underline{c}$ in a yield $r_i$ that would be high enough to increase the bond's value and simultaneously ensure winning. The shaded area in Figure \ref{figure_2.1} expresses bidders' willingness to participate in the auction and defines the boundaries for the  asset manager's payoff.
 Again, if the credit risk of the issuance is further reduced and $r^\ell=r^f$. The bidder will  tend to invest all the budget $c_i=\bar{c}$ at a rate equal to the risk-free rate, so he will not participate in the auction. In reverse, bidder $i$ will never bid for an interest rate $\Theta$ associated with a \enquote{high-yield} bond because it is beyond the upper bound $\bar{r}$.




\section{Existence of symmetric equilibrium} \label{Sec:Eq}

This section proves the existence of symmetric Bayesian Nash equilibrium. Under the assumption that all bidders choose the same strategy $b^*$, I examine the auction from the bidder's $i$ point of view. The analysis from other bidders' standpoints is symmetric. In a set of strategies $b^*_i(\cdot|\tau_{i}) $ of $n$ bidders, bidder $i$ maximizes the expected payoff for different realizations of signals. On this basis of  equilibrium, information updated through the signal space is the same and does not affect the outcome. Bidders privately observe the same signals before bid submission.

\begin{definition}
	For each strategy $b_i\in\mathcal{B}$, where $\mathcal{B}$ is the space of all strategies, there is an optimal strategy profile ${b^*}=(b_i^*,b_{-i}^*)$, which maximizes the expected payoff for all $i$, over the joint distribution $G(r^{s},\tau)$ and the signal space $\mathcal{S}$. That is, for pure strategies for bidder $i$:
	
	\begin{align*}		
		{\mathbb{E}}_{(r^{s},\tau)}[\pi_{i}(b_i^*,b_{-i}^*|s_i)]\geq{\mathbb{E}}_{(r^{s},\tau)}[\pi_{i}({b_i},b_{-i}^*|s_i)]		
	\end{align*}
	
\end{definition}

Bidders types are independent and identically distributed in a probability function that is common knowledge to everyone, and I assume that risk limits $r^{\ell^*}$ is symmetric and common to all.

I assume that $y=y^{n-1}$ is a random variable that attributes the type profiles $(n-1)$ remaining bidders, and $f_{y|\tau_i}$ denotes the conditional density function of $y$ given $\tau_i$. Bidder $i$  knows his type $\tau_i$ and that the highest value component-wise in $y$ is $\tau$.

The expected profit of bidder $i$ is given by:

\begin{equation}\label{exp_2.5}
	\mathbb{E}(\pi_i) = \displaystyle\int_{\mathbf{c^\ell}}^{\mathbf{\bar{c}}} 
	\alpha_i\bigg[    \hat{r}\big(b_i(\tau_i), b_{-i}(y)\big) -\mathbb{E}[r^{s}]                                                  \bigg] 
	f_{y|\tau_{i}} dy    
\end{equation}

where $\alpha_i$ is the allocation rule (Definition  \ref{definition3}), and 
$\mathbf{\bar{c}}=\max\limits_{j\in N/\{i\}}\bar{c}_j, \mathbf{c^\ell}=\max\limits_{j\in N/\{i\}} c^{\ell}_j$ respectively.

Let a minimum bid to participate in the auction define the risk characteristics of the bond. Herein, I parametrize the minimum bid by $\lambda\in (0,1)$. Hence, I assume that $b(c^\ell)=\lambda$ with an allocation equal to $\alpha(c^\ell)$.

Thus, the bidder's decision problem is to choose a bid $b$ to solve

\begin{equation*}
	\max\limits_{b^*}	{\mathbb{E}}\big[\pi_{i}(b_i^*,b_{-i}^*|s_i)\big],
\end{equation*}

if $b^*_i$ solves this problem, then the strategy $b^*_i$ is the best reply to $b_{-i}\dots b_n$. 

\begin{theorem} \label{Theorem1}
	The $n-tuple$ $(b^*, \dots, b^*)$ is a symmetric Nash equilibrium under uniform-price auctions when bidders follow the same bidding strategy concerning their budget and risk limits. For $\xi=\displaystyle\frac{\theta} {\Theta - \mathbb{E}[r^{s}] }$ and $\xi< \frac{1}{\lambda\,n}$ the bidding strategy is
	
	\begin{equation} \label{eq:6}
		b^*(c^*)=	\lambda \frac{\alpha(c^\ell)}{\alpha(c^*)} +\frac{1}{\xi \, n}\bigg[1-\frac{\alpha(c^\ell)}{\alpha(c^*)}\bigg]\ ,
	\end{equation} 
	with $c^*\in [c^\ell,\bar{c}]$.

\end{theorem}

\textit{Proof.} See the \ref{proof1}. \qed

Similar to a symmetric Cournot oligopoly, increasing the number of bidders lowers the equilibrium bidding strategy. 

The benchmark rate of \enquote{high-yield} bonds captured in parameter $\Theta$ is used to calculate a spread from the asset manager's expectation for the secondary market. If the asset manager expects the bond to be downgraded after the issuance, this spread becomes zero.
The parameter $\theta$ in the $\xi$ factor measures the response of stop-out yield to a bid's change. It attributes the oligopolistic effect of bidders upon the stop-out yield. As expected, the instructions set on the investment mandate directly impact the bidding strategy through the minimum bid $\lambda$ and allocation $\alpha(c^\ell)$. Additionally, the ratio of the minimum allocation $\alpha(c^\ell)$ to symmetric allocation $\alpha(c^*)$ affects the equilibrium bid. 


Next, I provide some basic comparative statics.

\begin{corollary}
	In symmetric equilibrium, the higher the oligopolistic power of bidders upon the stop-out yield, the lower the equilibrium bid.
\end{corollary}

\textit{Proof.} The result follows trivially since the equilibrium bid depends on inversely to $\theta$ (recall that $\theta$ appears in the numerator of $\xi$). \qed

Recall that the market power is the slope of inverse demand function \eqref{eq:4} and depends on the risk limit $r^\ell$ and the infimum budget that an investment manager is willing to invest in the issuance (Figure \ref{figure_2.1}). An asset manager with strong market power has low-risk acceptance ($r^\ell$) and may induce equilibrium to a lower stop-out yield.


\begin{corollary} \label{ref:2}
7
	In symmetric equilibrium, the stop-out yield is given by, 
	
	\[\hat{r}=\mathbb{E}[r^s]+({r}^{\ell^*}- \mathbb{E}[r^s])\frac{\alpha(c^\ell)}{\alpha(c^*)},  \]
	where ${r}^{\ell^*}$ is the symmetric risk limit for a symmetric minimum bid $b^*(c^\ell)=\lambda^*$.
\end{corollary}
\textit{Proof.}  See the Appendix \ref{Proof_Corollary_2.2}. \qed

Since bidding strategies respond to investors' signals,
the stop-out yield in the symmetric equilibrium reflects information about the bond's value concerning the risk limit imposed on the mandate.  The maximum spread an asset manager can earn from the resale in the secondary market seems to have a positive relationship. The value of the bond  from equation \eqref{exp_2.5} and Corollary \ref{ref:2} in the symmetry, equals to $({r}^{\ell^*}- \mathbb{E}[r^s])\frac{\alpha(c^\ell)}{\alpha(c^*)}$. It is easy to conclude that the risk limit is a boundary for the stop-out yield. An investment mandate with low-risk acceptance, ceteris paribus, moves $r^\ell$ downwardly, and the ratio $\frac{\alpha(c^\ell)}{\alpha (c^*)}$ declines (see Figure \ref{figure_2.1}). This can further decrease the stop-out yield in the symmetric equilibrium.

From corollary \ref{ref:2} and Figure \ref{figure_2.1}, the symmetric budget limit on the mandate, ceteris paribus, has a negative relation to the  symmetric stop-out yield. If bidders exhaust the available budget set on the investment mandate i.e., up to $\bar{c}$, the stop-out yield reaches the lower bound \underline{r}.


Proposition \ref{propo:1} shows that a high demand guided by narrow investment mandates would allocate the bond to more bidders since any bidder who participates in the issuance in the limit would bid the minimum.

\begin{proposition} \label{propo:1}
	In symmetric equilibrium, ceteris paribus, as the risk limit of bidders become strict (lower), i.e., $r^\ell$ goes to $r^f$, in the limit, the equilibrium bid equals $\lambda$.
\end{proposition}

\textit{Proof.} By the yield function, lower yields are associated with higher bids. I.e., for a decreasing sequence $(r^\ell_k)_{k \in N}$ corresponds to an increasing sequence $(c^\ell_k)_{k \in N}$. By letting $c^\ell$ to increase and given that $c^\ell<c^*$  then the ratio $\frac{\alpha(c^\ell)}{\alpha(c^*)}$ approaches to one. By equation (\ref{eq:6}) the result follows immediately. \qed


\section{Discussion} \label{discussion}

To interpret the equilibrium result following equation $\eqref{eq:4}$ I consider the following example. Suppose in the auction participate n=10 bidders with a symmetric market power $\theta=0.34$. The yield of the \enquote{high-yield} bonds in the market is $\Theta=8\%$, and the expectation for the secondary market $\mathbb{E}[r^s]=4\%$. Suppose the risk limit of the asset manager $i$ is up to $r^\ell=4.6\%$ for investing in an investment-grade bond. In that case, the infimum type revealed in the mechanism is ${c_i}^\ell=0.1$, and the mechanism will transform ${c_i}^\ell$ to a symmetric bid equal to $\lambda = 0.1$ (since the issuance in the model is normalized to one). The mechanism produdes a symmetric allocation equal the symmetric bid. So for $r^\ell$, the infinimum symmetric allocation will be $\alpha (c^\ell)=0.1$. If a bidder reveals in the mechanism a budget $c_i=0.169$ through the bidding strategy for an interest rate $r=2.95\%$. The allocation produced through the mechanism (equation $\ref{eq:4}$) for $c^*$ if $r=2.95\%$ was an equilibrium yield would be $\alpha(c^*)=0.148$. However, the symmetrical bid (equation \eqref{eq:6}) for an $r=2.95\%$ is $b(c^*)=0.0711$, and the residual supply of the issuance is 0.28. The symmetric equilibrium is reached for $b(c^*)=\lambda=0.1$, and the stop-out yield would be $\hat{r}=r^\ell=4.6\%$.

One can notice that the equilibrium bid is lower than the allocation the mechanism produces below the stop-out yield, i.e the symmetric risk limit. The first explanation is that the mechanism allocates lower yields first, as the auctioneer's revenue is increased with a low stop-out yield.  Conversely, the asset manager who participates in the issuance will exhaust the available budget by demanding a relatively high yield (close to the risk limit) to signal a high realization of the private information \citep{Bikhchandani1989}. The risk limit sets a boundary in demand, protecting the issuance from being underpriced which is a common issue in uniform auctions \cite{10.2307/3598014} and in the book-building process of corporate bonds \citep{10.2307/4494827}.  




\section{Conclusion}\label{Concl}

This paper introduces a mechanism to price corporate bonds in the primary market when a resale market exists, and bidders are restricted with an investment mandate for investment-grade bonds. In this setting, asset managers act as bidders who truthfully reveal their budget limit under a common knowledge risk limit. The model includes other features of normative importance, such as the interest rate of \enquote{junk bonds} and the market power of each bidder. 

I proved a symmetric Bayesian Nash equilibrium when winning bids and the issuance's yield are announced, provided that bids increase in the budget limit and the issuance's yield results from a linear rule. The equilibrium  is not unique, an expected result in uniform auctions\cite{Ausubel2014}. 

The risk limit set on the mandate works as a barrier to the bond's underpricing and positively affects the symmetric equilibrium yield. Investment mandates with low-risk acceptance provide stronger market power to an asset manager inducing the symmetric equilibrium bid downwardly. Similar to a symmetric Cournot oligopoly, the bidding is inversely affected by the number of competitive bidders. Another factor that affects the equilibrium bidding strategy is the spread between a benchmark rate in case of a \enquote{fire sale} and the expected yield in the secondary market, and it seems that the bidding strategy follows the same course with this spread.


%



Contrary to the current practice for pricing corporate bonds, the uniform-price auction is a well-understood rule by all parties and is a mechanism already used to price Treasury bills. Also, asset managers' widespread use of narrow investment mandates has implications in the primary market of corporate bonds. The analysis shows that even though asset managers may have adequate capital to invest in the corporate bond, the mandate confines it. As a result, the bids get closer to their values.

\appendix

\section{Proof of symmetric equilibrium}\label{proof1}

By Lemma \ref{lemma_2.1}, bidders directly reveal their types by announcing their budget limits. As I am examining the symmetric case, the allocation $\alpha_i$ that the mechanism produces is also symmetric. 
Thus,the expected profit from \eqref{exp_2.5} can be rewritten as:

\begin{align} 
	\mathbb{E}(\pi_i) &= \alpha_i \int_{c^\ell}^{ \bar{c}}  \hat{r}(b_i(c_i), b_{-i}(y)) f(y|c_i) dy  -\alpha_i\mathbb{E}[r^{s}] \int_{c^\ell}^{\bar{c}} f(y|c_i)  dy  \nonumber  \label{ep} \\ 
	&= \alpha_i \int_{ c^\ell}^{\bar{c}}  \hat{r}(b_i(c_i), b_{-i}(y)) f(y|c_i) dy -  \alpha_i\mathbb{E}[r^{s}]\big[F(\bar{c}|c_i) - F (c^\ell|c_i)\big]
\end{align}

I integrate by parts the integral $\displaystyle
\int_{c^\ell}^{\bar{c}}  \hat{r}(b_i(c_i), b_{-i}(y)) f(y|c_i) dy$, on the right hand side. By the continuity property of the distribution $F$, the probability of CDF for each bidder $i$ to bid a budget, $c_i\leq c_i^\ell$, equals zero.  Because none of the bidders will place a bid above their risk limit $r_i^\ell$ (Figure \ref{figure_2.1}).

This means that when $b_{-i}(c^\ell, r^\ell)=0$, the bond will not be issued, as none of the bidders can buy the whole issuance. In other words, the stop-out yield $\hat{r}(b_i(\tau_i), b_{-i}(c^\ell, r^\ell)) = \hat{r}(b_i(\tau_i), 0))=0$.

Substituting in equation (\ref{ep}), the optimization problem is:

\begin{align*} 
	\max_{c_i}	\mathbb{E}(\pi_i)  = \alpha_i \left[ \hat{r}(b_i(c_i), b_{-i}(\bar{c})) F({\bar{c}|c_i})  -  \int ^{\bar{c}}   _{ c^\ell}\hat{r}'(b_i(c_i), b_{-i}(y))   F(y|c_i) dy   -        \mathbb{E}[r^{s}] F(\bar{c}|c_i) \right]
\end{align*}

\begin{align*}
	\text{s.t.}
	\int^{ \bar{c}}_{c^\ell}\hat{r}'(b_i(\tau_i), b_{-i}(y))   F(y|c_i) dy  \leq 0
\end{align*}

To be a symmetric Bayesian Nash equilibrium, it is necessary that the first-order conditions to be zero. Ex ante, at the optimum, the expected stop-out yield can not be further diminished, hence the aforementioned constraint is satisfied with the equality.


Because of the symmetry, all bidders share the same type $c^*$. By Lemma \ref{lemma_2.1} bidders maximize with respect to their budget limit.

\begin{align}
	0 &=\frac{\partial \mathbb{E}(\pi_i)}{\partial c_i}
	\Bigr\rvert_{(c_i=c^*)} \nonumber \\
	&=\Big(\big[ \hat{r}(b_i(c_i), b_{-i}(\bar{c})) F({\bar{c}|c_i})    
	-        \mathbb{E}[r^{s}] F(\bar{c}|c_i)              \big]\alpha_{i}(c_i)\Big)'  \nonumber \\
	&=\alpha'_{i}(c^*) \hat{r}(b_{i}(c^*), b_{{-i}}(\bar{c}=c^*)) F(\bar{c}=c^*|c^*) - \alpha'_{i}(c^*) \mathbb{E}[r^{s}]F(\bar{c}=c^*|c^*) \nonumber\\
	& \bigr. + \alpha_{i}(c^*) \hat{r}'(b_{i}(c^*), b_{-i}(\bar{c}=c^*))  F(\bar{c}=c^*|c^*)  \label{eq:9} 
\end{align}

I substitute \eqref{eq:4} to \eqref{eq:9} and for simplicity reasons, I denote  $\rho(c^*)=\frac{\alpha'_{i}(c^*) }{\alpha_{i}(c^*)}$, which is the relative rate of change for the symmetric allocation $\alpha_i=\alpha^*$, and with $b^*$ the symmetric bidding strategy. Thus, 

\begin{align*}
	\rho(c^*)\, [\Theta - n \, \theta \, {b^*}(c^*)] + [-\theta \, n \, {b^*}^{\prime}(c^*)] \,   -\rho(c^*) \, \mathbb{E}[r^{s}]=0 
\end{align*}

Denominating with $(-\theta n )$  and by substitution of $\xi=\displaystyle\frac{\theta} {\Theta - \mathbb{E}[r^{s}] }$, where  $\xi< \frac{1}{\lambda\,n}$, I result to a first-order non-homogeneous differential equation:

\begin{equation*}
	{b^*}^{\prime}(c^*)  +\rho(c^*) \, {b^*}(c^*)=\frac{\rho(c^*) }{\xi\, n } \label{eq:10}
\end{equation*}

The solution to the first-order differential equation is given by

\begin{align*} \label{eq:11}
	{b^*}(c^*)&=\displaystyle \mathrm{e}^{- \int \rho (c^*) \, dc^*} \bigg(\displaystyle\int \mathrm{e}^{\int \rho(c^*) \, dc^*}\, \frac{\rho(c^*) \,} {\xi\, n } \,  \, dc^* \, + \, \Gamma \bigg)\\ \nonumber &= \displaystyle \mathrm{e}^{\ln\alpha^{-1}(c^*)} \bigg(\displaystyle\int \mathrm{e}^{\ln\alpha(c^*)}  \frac{\rho(c^*) \,} {\xi\, n } \,  \, dc^* \, + \, \Gamma \bigg)\\ \nonumber
	&=\frac{1}{\alpha(c^*)}\bigg(\displaystyle\int  \frac{\alpha'(c^*) \,} {\xi\, n } \,  \, dc^* \, + \, \Gamma \bigg) \\ \nonumber
	&=\frac{1}{\alpha(c^*)}\bigg(\frac{\alpha(c^*) \,} {\xi\, n }\, + \, \Gamma\bigg)\nonumber 
\end{align*}
where $\Gamma$ is an arbitrary constant. Thus, I conclude
\begin{equation}\label{equation_2.10}
	b^*(c^*)=\frac{1}{\xi n} + \frac{\Gamma}{a(c^*)},\, \text{with} \, c^*\in[c^\ell,\bar{c}]
\end{equation}

Now since $b(c^\ell)=\lambda$ is the initial condition of the differential equation, then the value of the constant $\Gamma=\alpha(c^\ell)[\lambda-\frac{1}{\xi \, n}]$, where $\alpha(c^\ell)\in(0,1)$ and corresponds to the minimum allocation of the winning bidder. Thus, the solution of equation \eqref{equation_2.10} is unique and can be re-written:

\begin{align*}
	b^*(c^*)&=\frac{1}{\xi\, n} + \frac{\alpha(c^\ell)[\lambda-\frac{1}{\xi \, n}]}{a(c^*)} \\ \nonumber
	&=\frac{1}{\xi\, n} + \frac{\alpha(c^\ell)\,\lambda}{\alpha(c^*)}-\frac{\alpha(c^\ell)}{\alpha(c^*)\,\xi \, n} \\ \nonumber
	&=\lambda \frac{\alpha(c^\ell)}{\alpha(c^*)} +\frac{1}{\xi \, n}\bigg[1-\frac{\alpha(c^\ell)}{\alpha(c^*)}\bigg].
\end{align*} 

Bidders maximize a linear expected payoff function under a linear constraint. Hence, it is anticipated that the second derivative is zero with respect to the strategic variable, that by direct revelation mechanism is defined to be the budget limit $c_i$. Below, it is illustrated that the second-order derivative becomes zero when I substitute the equilibrium bid:

\begin{equation*}
	\begin{split}
	\frac{\partial^2\mathbb{E}(\pi_i)}{\partial c^2_i}	\Bigr\rvert_{(c_i=c^*)}  &=	\alpha''_{i}(c^*) \hat{r}(b_{i}(c^*), b_{{-i}}(\bar{c}=c^*)) F(\bar{c}=c^*|c^*)  \\
	 &\quad + 2\alpha'_{i}(c^*) \hat{r}'(b_{i}(c^*), b_{-i}(\bar{c}=c^*))  F(\bar{c}=c^*|c^*)  \\
	 &\quad + \alpha_{i}(c^*) \hat{r}''(b_{i}(c^*), b_{-i}(\bar{c}=c^*))  F(\bar{c}=c^*|c^*) \\
	 &\quad - \alpha''_{i}(c^*) \mathbb{E}[r^{s}]F(\bar{c}=c^*|c^*)   
	\end{split}
\end{equation*}

Because the allocation rule is a linear increasing and differential function on the budget limit  $c$, I have $\alpha''_{i}(c^*)=0$ and by substituting in (4)
 \[=-2\alpha'_{i}(c^*) \,n\, \theta\, {b^*}^{\prime}(c^*) \,  F(\bar{c}=c^*|c^*) - \alpha_{i}(c^*) \,n\, \theta\, {b^*}^{\prime \prime}(c^*) \,  F(\bar{c}=c^*|c^*)\]

Replacing the first and the second derivative of equation (6)  

\begin{equation*}\label{ch_2_equation_7}
	\begin{split}
		&=2 \rho ^2 (c^*)\theta\,n\,\lambda \, \alpha (c^\ell) F(c^*|c^*) - 2 \rho ^2 (c^*)\, \big(\Theta - \mathbb{E}[r^s]\big)\,\alpha (c^\ell) F(c^*|c^*) \\ 
		 &\quad - 2 \rho ^2 (c^*)\theta\,n\,\lambda \, \alpha (c^\ell) F(c^*|c^*)   + 2 \rho ^2 (c^*)\, \big(\Theta - \mathbb{E}[r^s]\big)\,\alpha (c^\ell) F(c^*|c^*) =0 
	\end{split}
\end{equation*}
It is a standard result in uniform auctions that there is no unique equilibrium \citep{Ausubel2014}. Though a recent paper shows that under certain conditions it exists uniqueness \citep{Burkett2020}. 
\qed

\section{Proof of Corollary 2.2} \label{Proof_Corollary_2.2}

In the symmetric case, the symmetric equilibrium yield from \eqref{eq:4} is: 

\[\hat{r}= \Theta - \theta\, n \, b^*(c^*)\]

Substituting equation (\ref{eq:6}) in (\ref{eq:4}) 
I can rewrite equivalently:
\begin{align*}
	\hat{r} &= \Theta - \theta n \bigg(\lambda \frac{a(c^\ell)}{a(c^*)} + \frac{1}{\xi n}\bigg[1-\frac{a(c^\ell)}{a(c^*)}\bigg]\bigg) \\
	&=\Theta - \frac{\theta \,n \,\lambda \,\alpha(c^\ell) }{\alpha (c^*)} - \Theta + \mathbb{E}[r^s] +(\Theta - \mathbb{E}[r^s]) \frac{\alpha(c^\ell)}{\alpha (c^*)}\\
	&=\mathbb{E}[r^s] + \big(\Theta -\theta\,n\,\lambda - \mathbb{E}[r^s]\big)\frac{\alpha(c^\ell)}{\alpha (c^*)}\\
	&=\mathbb{E}[r^s] + \big(\Theta -\theta\,n\,b^*(c^\ell)- \mathbb{E}[r^s] \big)\frac{\alpha(c^\ell)}{\alpha (c^*)}
\end{align*}
By substituting with the symmetric minimum bid $b^*(c^\ell)=\lambda$, I result in an equilibrium yield equal to the symmetric risk limit that is ${r}^{\ell^*}=\Theta -\theta\,n\,b^*(c^\ell)$. Thus, I conclude that

\[\hat{r}=\mathbb{E}[r^s] + \big({r}^{\ell^*}- \mathbb{E}[r^s] \big)\frac{\alpha(c^\ell)}{\alpha (c^*)}.\]

\qed


\bibliographystyle{elsarticle-harv}

\bibliography{Bibliography}







\end{document}